\documentclass[conference]{IEEEtran}
\IEEEoverridecommandlockouts
\usepackage{amsmath,amssymb,physics,tabularx,graphicx,multirow,hhline,algorithm,mathtools,algpseudocode,varwidth,mathrsfs}
\allowdisplaybreaks
\IEEEoverridecommandlockouts
\ifCLASSINFOpdf
\else
\fi

\ifCLASSOPTIONcompsoc \usepackage[caption=false,font=normalsize,labelfont=sf,textfont=sf]{subfig}
\else
\usepackage[caption=false,font=footnotesize]{subfig}
\fi
\ifCLASSINFOpdf
  
\else
  
\fi

\begin{document}
%

\title{Detection of Cyber Attacks in Renewable-rich Microgrids Using Dynamic Watermarking}


%
\author{\IEEEauthorblockN{Tong Huang,
Bin Wang,
Jorge Ramos-Ruiz, Prasad Enjeti, P. R. Kumar, and
Le Xie}
\IEEEauthorblockA{Department of Electrical and Computer Engineering\\
Texas A\&M University, College Station, Texas
\\ Email:\{\texttt{tonghuang}, \texttt{binwang}, \texttt{jorgeramos}, \texttt{enjeti}, \texttt{prk}, \texttt{le.xie}\}\texttt{@tamu.edu}}

\thanks{This work is supported in part by NSF Science \& Technology Center Grant CCF-0939370, and the Power Systems Engineering Research
Center.}}


\maketitle

\begin{abstract}
This paper presents the first demonstration of active defense for defending renewable-rich microgrids against cyber attacks. Cyber vulnerabilities in such microgrid setting are identified. A defense mechanism based on the dynamic watermarking is proposed for detecting cyber anomalies in microgrids. The proposed mechanism is shown to be readily implementable and it has theoretically provable performance in term of detecting cyber attacks. The effectiveness of the proposed mechanism is tested and validated in a Texas A\&M 4-bus microgrid testbed with 100\% solar penetration.
\end{abstract}


%
\IEEEpeerreviewmaketitle

\section{Introduction}
Renewable energy sources (RESs) in distribution systems, e.g. rooftop photo-voltaic (PV) panels and micro-turbines, provide a key solution to mitigate potential climate change. Microgrids, namely, small but autonomous power systems, are a crucial concept for integrating and managing these small-scale RESs. In a microgrid, power electronic inverters serve as interfaces between network and distributed RESs associated with energy storage systems \cite{4118327}. Compared with large-scale power grids with giant generators, the inverter-based microgrids lack physical inertia, resulting in potentially large voltage and frequency deviation when subject to disturbances \cite{4118327,1626398}

In order to mitigate the impact caused by an insufficient inertia in inverter-based microgrids, a feedback loop, i.e. a droop controller, is embedded into the inverter of each distributed generation unit (DGU) to emulate governor's behavior of a conventional synchronous generator. Such droop controllers constitute the primary control of an inverter-based microgrid, whose control strategies can be categorized into PQ inverter control and voltage source inverter (VSI) control \cite{1626398}. Besides, the droop control enables peer-to-peer energy transaction control in microgrid systems \cite{8598671}. A popular choice of control strategy is to regulate the terminal behavior of an inverter, i.e. real and reactive power for PQ inverter control and voltage magnitude and frequency for VSI control, via tuning setpoints of the inverter \cite{1626398}. 

With the proliferation of new DGUs, the concomitant control and information infrastructure become new elements of the power grid that are exposed to cyber attacks. There arises vulnerabilities where adversaries can compromise an individual DGU or even the entire microgrid via maliciously manipulating the terminal measurements fed into droop controllers. The risk of cyber attacks in transmission systems, such as system/device-level instabilities and increase of operational costs \cite{8345676,Liu:2009:FDI:1653662.1653666}, also exists in renewable-rich microgrids. Reference \cite{8260848} investigates the negative impact of false data injection attacks on microgrids from a stability perspective. As a deep penetration of renewables in distribution systems leads to an intensive deployment of inverters in microgrids, it is imperative to equip DGU inverters with a functionality for detecting cyber anomalies.

In this paper, we present the first demonstration of using an active mechanism for defending renewable-rich microgrids against cyber attacks on the measurements fed into droop controllers. We illustrate the possibility that a malicious distortion of limited number of sensors can incur a system-level instability in microgrids. We design the defense mechanism by tailoring the dynamic watermarking technique \cite{7738534,7945354} to the microgrid context, for the purpose of detecting cyber attacks on the sensors used by droop controllers. Advantages of the proposed mechanism contain two aspects: 1) the defense mechanism is practically implementable, as it does not require any hardware upgrade on inverters; 2) it is theoretically guaranteed that the watermarking-based mechanism can detect \emph{any} malicious manipulations on measurements fed into droop controllers, even if attackers possess extensive knowledge of physical and statistic models of microgrids.

The rest of this paper is organized as follows: Section \ref{sec:mathematic_description_of_a_microgrid} describes the dynamics of a microgrid with droop control and identifies the cyber vulnerability in such a system; Section \ref{sec:cyber_attack_detection_based_dynamic_watermarking} introduces the defense mechanism based on dynamic watermarking; Section \ref{sec:case_study} tests the proposed defense mechanism in a synthetic renewable-rich microgrid; and Section \ref{sec:conclusion} concludes this paper and points out future research directions.

\section{Problem Formulation} 
\label{sec:mathematic_description_of_a_microgrid}
This section aims to characterize dynamics of a microgrid with droop control and to define cyber attacks in a microgrid. We first present differential and algebraic equations (DAEs) for each component in a microgrid. Then, a system-level small-signal model for the microgrid will be derived. Finally, we identify cyber vulnerabilities in the microgrid with prevailing frequency/voltage droop controllers.

\subsection{Dynamics of a Renewable-Rich Microgrid} 
\label{sub:dynamics_of_a_microgrid}
\subsubsection{Modeling of Network and Loads}
Given a microgrid with $b$ buses, $n$ DGUs and $m$ loads, denote the sets of all buses, DGU-connected buses and load buses respectively as $\mathcal{B}$, $\mathcal{D}$ and $\mathcal{L}$. The constraints resulting from the microgrid network are described by power flow equations 
\begin{equation}\label{eq:network}
    S_{i}- V_{i}\sum_{j\in \mathcal{B}} V_{j}(G_{ij}-jB_{ij})e^{j\theta_{ij}}=0, \text{ } i\in \mathcal{B}
\end{equation}
where $S_{i}=P_{i} + jQ_{i}$ and $V_{i}$ are respectively the power injection and voltage magnitude at bus $i$; and $\theta_{ij}$, $G_{ij}$ and $B_{ij}$ represent the angle difference, conductance and susceptance of the branch between buses $i$ and $j$. According to different bus types, the power injection in \eqref{eq:network} can be categorized into the following four cases
\begin{equation}
    S_{i}=
		    \begin{cases} 
            S_{\text{DGU}ni} & n_{i}\in \mathcal{D}, m_{i}\not\in \mathcal{L}\\
            -S_{\text{L}mi} & n_{i}\not\in \mathcal{D}, m_{i}\in \mathcal{L} \\
            S_{\text{DGU}ni}-S_{\text{L}mi} & n_{i}\in \mathcal{D}, m_{i}\in \mathcal{L} \\
            0 & \text{Otherwise}
            \end{cases}
\end{equation}
where $S_{\text{DGU}ni} = P_{\text{DGU}ni}+jQ_{\text{DGU}ni}$ is the power output of $n_{i}$-th DGU, which is connected to bus $i$; and $S_{\text{L}mi} = P_{\text{L}mi}+jQ_{\text{L}mi}$ is the power consumption of $m_{i}$-th load, which is connected to bus $i$. For simplicity, all loads are represented by constant power during dynamics.
\subsubsection{Modeling of DGU}
\begin{figure}[htb]
	\centering
	\includegraphics[width = 2.6in]{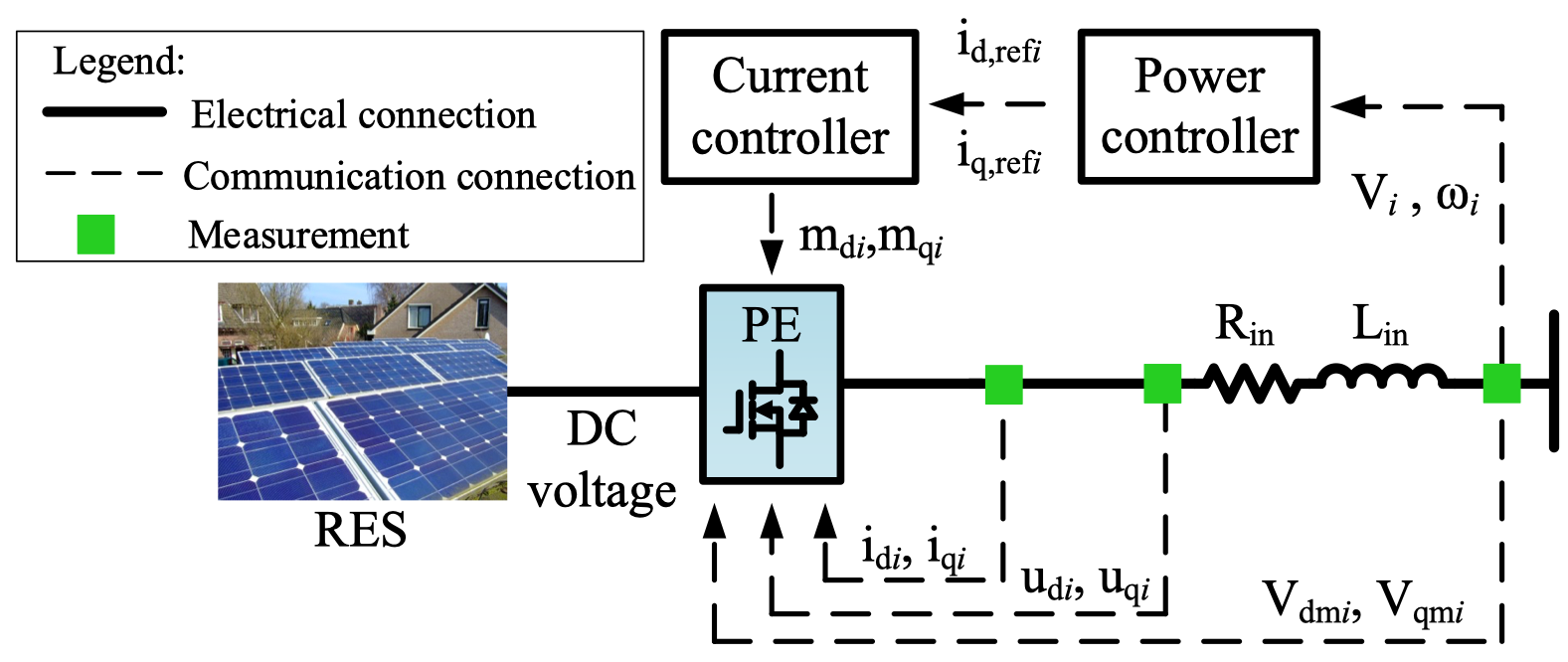}
	\caption{DGU inverter block diagram.}
	\label{fig:diag}
\end{figure}

This paper focuses on the frequency droop control of DGU, therefore, the dynamics associated with the primary source and the DC link are ignored for simplicity, where an ideal DC voltage source with a voltage at $V_{\text{dc}}$ is adopted to represent the DC link. The block diagram is shown in Fig. \ref{fig:diag}.

The input signals of a power controller are the voltage magnitude $V_i$ and frequency $\omega_i$. First-order systems are adopted to mimic measurement delays as shown below
\begin{subequations} \label{eq:mea}
	\begin{align}
		\dot{\omega}_{\text{m}i}=(\omega_{i} - \omega_{\text{m}i})/T_{\omega},\\
		\dot{V}_{\text{m}i}=(V_{i} - V_{\text{m}i})/T_{V},\\
		\dot{\theta}_{\text{m}i}=(\theta_{i} - \theta_{\text{m}i})/T_{\theta},
	\end{align}
\end{subequations}
where $\omega_{\text{m}i}$ and $V_{\text{m}i}$ are the measured frequency and voltage magnitude at bus $i$, respectively; $\theta_i$ and $\theta_{\text{m}i}$ are the voltage phase angle and its measured version at bus $i$; and $T_{\omega}$, $T_V$, and $T_{\theta}$ are time constants.

The power controller uses pre-defined droop characteristics to determine the setpoints of active and reactive power, based on the following control laws:
\begin{subequations} \label{eq:droop}
	\begin{align}
		P_{\text{ref}i} &= \alpha_{\text{p}i}(\omega_{\text{m}i} - \omega_{\text{ref}i})+ \beta_{\text{p}i} \int_0^{t}(\omega_{\text{m}i}(\tau)-\omega_{\text{ref}i})\text{d}\tau + P_{\text{ref}i}^*,\\
		Q_{\text{ref}i} &= \alpha_{\text{q}i}(V_{\text{m}i} - V_{\text{ref}i})+ \beta_{\text{q}i} \int_0^{t}(V_{\text{m}i}(\tau)-V_{\text{ref}i})\text{d}\tau + Q_{\text{ref}i}^*,
	\end{align}
\end{subequations}
where $P_{\text{ref}i}^*$ and $Q_{\text{ref}i}^*$ are steady-state real and reactive power; setpoints $\omega_{\text{ref}i}$ and $V_{\text{ref}i}$ are specified by secondary control of the microgrid; and scalars $\alpha_{\text{p}i}$, $\beta_{\text{p}i}$, $\alpha_{\text{q}i}$, and $\beta_{\text{q}i}$ are droop control parameters of DGU connected to bus $i$. Then, the power controller determines the reference signals ($i_{\text{d,ref}i}$ and $i_{\text{q,ref}i}$) for a current controller according to
\begin{subequations} \label{eq:cur}
	\begin{align}
		i_{\text{d,ref}i}=(P_{\text{ref}i}V_{\text{dm}i}+Q_{\text{ref}i}V_{\text{qm}i})/V_{\text{m}i},\\
		i_{\text{q,ref}i}=(P_{\text{ref}i}V_{\text{qm}i}-Q_{\text{ref}i}V_{\text{dm}i})/V_{\text{m}i},
	\end{align}
\end{subequations}
where $V_{\text{dm}i}$, $V_{\text{qm}i}$ are the d-, and q-axis components of the measured terminal voltage $V_{\text{m}i}$ at bus $i$.

The current controller generates the voltage modulation signals for a DGU inverter, i.e. $m_{\text{d}i}$ and $m_{\text{q}i}$, based on the current reference signals from the power controller. The dynamics of the current controller is described by the following DAEs:
\begin{subequations} \label{eq:cc_ae}
	\begin{align}
	    	&\dot{\gamma}_{\text{d}i}=i_{\text{d,ref}i} - i_{\text{d}i}, \quad \dot{\gamma}_{\text{q}i}=i_{\text{q,ref}i} - i_{\text{q}i},\label{eq:gamma_dynamics}\\
		&m_{\text{d}i} = K_{\text{p1}i}(i_{\text{d,ref}i} - i_{\text{d}i}) + K_{\text{i1}i}\gamma_{\text{d}i},\\
		&m_{\text{q}i} = K_{\text{p2}i}(i_{\text{q,ref}i} - i_{\text{q}i}) + K_{\text{i2}i}\gamma_{\text{q}i},
	\end{align}
\end{subequations}
where $K_{\text{p}1i}$, $K_{\text{i}1i}$, $K_{\text{p}2i}$, and $K_{\text{i}2i}$ are control parameters.

Based on the voltage modulation signals, the output of the inverter is determined by
\begin{equation} \label{eq:inv}
		u_{\text{d}i}=m_{\text{d}i}V_{\text{dc}i}, \quad u_{\text{q}i}=m_{\text{q}i}V_{\text{dc}i}.
\end{equation}
The output current dynamics of the inverter are described by
\begin{subequations} \label{eq:outputi}
	\begin{align}
		\dot{i}_{\text{d}i}=(V_{\text{d}i} +R_{\text{in}i}i_{\text{d}i} -u_{\text{d}i} - \omega_{i}L_{\text{in}i}i_{\text{q}i})/L_{\text{in}i},\\
	\dot{i}_{\text{q}i}=(V_{\text{q}i} +R_{\text{in}i}i_{\text{q}i} -u_{\text{q}i} - \omega_{i}L_{\text{in}i}i_{\text{d}i})/L_{\text{in}i}.
	\end{align}
\end{subequations}

\subsubsection{Compact Form for Microgrid Model}
In order to obtain compact expressions of microgrid dynamics, we first define some vectors as follows:
\begin{subequations}\label{eq:notation}
	\begin{align}
		&\mathbf{u}_{\text{ref}i}:= [P_{\text{ref}i},Q_{\text{ref}i}]^{\top}, \quad \mathbf{y}_i:= [\omega_i, V_i]^{\top},\\
		&\mathbf{u}_{\text{ref}} = [\mathbf{u}_{\text{ref}1}^{\top}, \ldots, \mathbf{u}_{\text{ref}i}^{\top},\ldots, \mathbf{u}_{\text{ref}n}^{\top}]^{\top},\\
		&\mathbf{u}_{\text{L}}:= [P_{\text{L}1},Q_{\text{L}1}, \ldots, P_{\text{L}i}, Q_{\text{L}i}, \ldots, P_{\text{L}m},Q_{\text{L}m}]^{\top},
	\end{align}
\end{subequations}
where $P_{\text{ref}i}$ and $Q_{\text{ref}i}$ are the real and reactive power, respectively; and $P_{\text{L}i}$ and $Q_{\text{L}i}$ are the real and reactive power of the $i$-th load. The dynamics of a microgrid without droop controllers can be described by
\begin{subequations}\label{eq:nonlinear}
	\begin{align}
		&\dot{\mathbf{x}} = \mathbf{f}(\mathbf{x}, \mathbf{y}, \mathbf{u}), \\
		&\mathbf{g}(\mathbf{y}, \mathbf{u}) = 0,
	\end{align}
\end{subequations}
where the state vector $\mathbf{x}$ comprises all 7 states of $n$ DGUs involved in \eqref{eq:mea}, \eqref{eq:gamma_dynamics} and \eqref{eq:outputi}; vector $\mathbf{u}=[\mathbf{u}_{\text{ref}}^{\top}, \mathbf{u}_{\text{L}}^{\top}]^{\top}$; and bus frequencies and other variables constitute output vector $\mathbf{y}$.


\subsection{Small-Signal Model of a Renewable-Rich Microgrid} 
\label{sub:small_signal_model_of_a_microgrid}
Define vector $\mathbf{y}_0$ by collecting DGU terminal voltage magnitudes $V_i$ and frequencies $\omega_i$ for all $i = 1, 2,\ldots, n$, i.e., $\mathbf{y}_0 := [\mathbf{y}_1^{\top}, \ldots, \mathbf{y}_n^{\top}]^{\top}$. 
Next, we derive the small-signal model of a microgrid.
\subsubsection{Linearized Model of a Microgrid Without Droop Control} 
\label{ssub:linearized_model_of_a_microgrid_}
The dynamical behavior of system \eqref{eq:nonlinear} around its equilibrium point can be approximated by
\begin{subequations}\label{eq:open_loop_small_signal}
 	\begin{align}
 		&\Delta \dot{\mathbf{x}} = A\Delta\mathbf{x}+B_{\text{ref}} \Delta \mathbf u_{\text{ref}}+B_{\text{L}}\Delta \mathbf u_{\text{L}} + \boldsymbol{\xi}',\\
 		&\Delta \mathbf{y}_0 = C \Delta\mathbf{x} + \boldsymbol{\gamma}',
 	\end{align}
 \end{subequations} 
where $\Delta \mathbf{x}$, $\Delta \mathbf{u}_{\text{ref}}$, $\Delta \mathbf u_{\text{L}}$, and $\Delta \mathbf{y}_0$ are \emph{deviation} of states $\mathbf{x}$, setpoint inputs $\mathbf{u}_{\text{ref}}$, load inputs $\mathbf{u}_{\text{L}}$, and truncated outputs $\mathbf{y}_0$ from their corresponding values at the equilibrium point, respectively; matrices $A$, $B_{\text{ref}}$, $B_{\text{L}}$ and $C$ can be obtained by using analytical/numerical linearization techniques \cite{kundur1994power,8116598,8362302,huang2018forced}; and $\boldsymbol{\xi}'\sim \mathcal{N}(0, R')$ and $\boldsymbol{\gamma}'\sim \mathcal{N}(0,V')$ are process noise and measurement noise, respectively \cite{8345676}.
\subsubsection{Linearized Model of A Microgrid with Droop Control} 
\label{ssub:linearized_droop_control_policies}
The small-signal model for the control policies \eqref{eq:droop} is
\begin{subequations} \label{eq:small_signal_droop}
	\begin{align}
		\Delta P_{\text{ref}i} &= \alpha_{\text{p}i} \Delta \omega_i+ \beta_{\text{p}i} \int_0^{t}\Delta\omega_i(\tau)\text{d}\tau,\\
		\Delta Q_{\text{ref}i} &= \alpha_{\text{q}i} \Delta V_i+ \beta_{\text{q}i} \int_0^{t}\Delta V_i(\tau)\text{d}\tau,
	\end{align}
\end{subequations}
where $\Delta P_{\text{ref}i} = P_{\text{ref}i}-P_{\text{ref}i}^*$, $\Delta Q_{\text{ref}i} = Q_{\text{ref}i}-Q_{\text{ref}i}^*$, $\Delta\omega_i(\tau)=\omega_i(\tau)-\omega_{\text{ref}i}$, and $\Delta V_i(\tau)=V_i(\tau)-V_{\text{ref}i}$.
The control policies \eqref{eq:small_signal_droop} for all DGUs can be expressed in a compact form:
\begin{equation}
	\Delta\mathbf{u}_{\text{ref}i} = \mathbf{h}_i(\Delta\mathbf{y}_i), \quad \forall i = 1, 2, \ldots,n,
	\label{eq:droop_compact}
\end{equation}
where $\mathbf{h}_i(\cdot)$ denotes local droop control policies in \eqref{eq:small_signal_droop} in the $i$-th DGU. The closed-loop model for $n$ interconnected DGUs with droop controllers are given by \eqref{eq:open_loop_small_signal} and \eqref{eq:droop_compact}. Note that the secondary and tertiary control of the microgrid are ignored.

In practice, the control law \eqref{eq:droop} is implemented discretely based on the sampling rate $\Delta T_{\text{s}}$ of the sensors measuring $\mathbf{y}_i$. Based on the zero-order hold discretization method, the discrete version of \eqref{eq:small_signal_droop} is
\begin{subequations} \label{eq:discrete_small_signal_droop}
	\begin{align}
		\Delta P_{\text{ref}i}[k] &= \alpha_{\text{p}i} \Delta \omega_i[k]+ \beta_{\text{p}i} \Delta T_{\text{s}} \sum_{\kappa = 1}^k\Delta\omega_i[\kappa],\\
		\Delta Q_{\text{ref}i}[k] &= \alpha_{\text{p}i} \Delta V_i[k]+ \beta_{\text{p}i} \Delta T_{\text{s}} \sum_{\kappa = 1}^k\Delta V_i[\kappa].
	\end{align}
\end{subequations}
Equation \eqref{eq:discrete_small_signal_droop} can be expressed in the following compact form:
\begin{equation}\label{eq:discrete_droop_compact}
	\Delta\mathbf{u}_{\text{ref}i}[k] = \mathbf{h}_{\text{d}i}\left(\Delta\mathbf{y}_i^{k}\right), \quad \forall i = 1, 2, \ldots,n,
\end{equation}
where $\Delta\mathbf{y}_i^{k} := \left\{\Delta\mathbf{y}_i[1], \Delta\mathbf{y}_i[2], \ldots, \Delta\mathbf{y}_i[k]\right\}$, which contains all $k$ measurement samples fed into the droop controller at DGU $i$. Similarly, the discrete version of \eqref{eq:open_loop_small_signal} is
\begin{subequations}\label{eq:discrete_open_loop_small_signal}
 	\begin{align}
 		& \begin{aligned}
 			\Delta \mathbf{x}[k+1] = &A_{\text{d}}\Delta\mathbf{x}[k]+B_{\text{refd}} \Delta \mathbf u_{\text{ref}}[k]\\&+B_{\text{Ld}}\Delta \mathbf u_{\text{L}}[k] + \boldsymbol{\xi}'[k+1],
 		\end{aligned}\\
 		&\Delta \mathbf{y}_0[k+1] = C_{\text{d}}\Delta\mathbf{x}[k+1] + \boldsymbol{\gamma}'[k+1],
 	\end{align}
 \end{subequations}
 where matrices $A_{\text{d}}$, $B_{\text{refd}}$, $B_{\text{Ld}}$ and $C_{\text{d}}$ are obtained by standard discretization methods, such as the zero-order hold method, or Tustin method. Note that the state-space model can be identified from measurements via some system identification techniques, if the physical model of microgrids is not available.

\subsection{Cyber Vulnerabilities in Microgrids With Droop Control} 
\label{sub:cyber_vulneral}
In the microgrid regulated by droop controllers, there arises vulnerabilities where adversaries can compromise the microgrid by manipulating the edge measurements $\Delta\mathbf{y}_i^k$. When there is no cyber attack, the frequency and voltage sensors honestly report actual measurements to droop controllers. However, under cyber attacks, the frequency and voltage sensors are forced to stream a different vector $\Delta\mathbf{z}_i^k$ to the droop controllers. In such a case, control commands $\Delta P_i$ and $\Delta Q_i$ are computed based on the reported vector $\Delta\mathbf{z}_i^k$, rather than actual measurements $\Delta\mathbf{y}_i^k$. As a result, both performance and safety of the microgrid may be compromised. Based on the above attack strategy, several attack templates, e.g., noise injection attack, replay attack, and destabilization attack, are proposed in \cite{8345676}. 


\section{Cyber Attack Detection via Dynamic Watermarking} 
\label{sec:cyber_attack_detection_based_dynamic_watermarking}
In this section, we apply the dynamic watermarking technique \cite{7738534,7945354} to detect cyber anomalies in the microgrid with droop controllers. We first illustrate how to inject ``watermark'' signals to the microgrid. Then, cyber attack indicators are designed based on dynamic watermarking. Finally, built upon the proposed indicators, a threshold test for detecting cyber attacks is presented.

\subsection{Injection of Watermark Signal} 
\label{sub:injection_of_watermark_signal}
For DGU $i$, instead of directly applying control commands $\mathbf{h}_i(\Delta\mathbf{z}_i^k)$ to change $\Delta P_{\text{ref}i}$ and $\Delta Q_{\text{ref}i}$, the droop controller of DGU $i$ superposes an independent and identically distributed (i.i.d.) watermark signal $\mathbf{e}_i [k]$ upon the control command, namely, 
\begin{equation} \label{eq:private_inj}
	\Delta\mathbf{u}_{\text{ref}i}[k] = \mathbf{h}_{\text{d}i}\left(\Delta\mathbf{y}_i^{k}\right)+ \mathbf{e}_i [k],
\end{equation}
where $\mathbf{e}_i[k] \sim \mathcal{N}(0, \nu_e I_{2\times 2})$ and $I_{2\times 2}$ denotes a $2$-by-$2$ identity matrix. Note that the realization of the watermark signal $\mathbf{e}_i [k]$ is only known by the droop controller $i$. With such a secret watermark signal, certain statistical properties related to the watermark signal are expected to exhibit in the actual measurements $\Delta \mathbf{y}_i^k$, while these properties disappear in excessively distorted measurements \cite{8345676}. Hence, by checking the existence of these statistic properties, cyber attacks on measurements fed into droop controllers can be detected. In the following we will design two indicators for checking the existence of these statistic properties.



\subsection{Cyber Attack Indicators} 
\label{sub:Cyber Attack Indicators}
For the droop controller at DGU $i$, the rest of the microgrid seen by this droop controller is an open-loop system with inputs $\{\Delta \mathbf{u}_{\text{ref}i}, \Delta \mathbf{u}_{\text{L}}\}$ and outputs $\Delta \mathbf{y}_i$. The input-output dynamics of such an open-loop system are equivalent to that of the following reduced-order model, viz.,
\begin{subequations}\label{eq:minimal_realization}
	\begin{align}
		& \begin{aligned}
 			\Delta \mathbf{x}_i[k+1] = &A_{\text{m}i}\Delta\mathbf{x}_i[k]+B_{\text{refm}i} \Delta \mathbf u_{\text{ref}i}[k]\\&+B_{\text{Lm}i}\Delta \mathbf u_{\text{L}}[k] + \boldsymbol{\xi}[k+1],
 		\end{aligned}\\
 		&\Delta \mathbf{y}_i[k+1] = C_{\text{m}i}\Delta\mathbf{x}_i[k+1] + \boldsymbol{\gamma}[k+1],
	\end{align}
\end{subequations}
where $A_{\text{m}i}$, $B_{\text{refm}i}$, $B_{\text{Lm}i}$, and $C_{\text{m}i}$ are obtained by minimally realizing \cite{8345676} the system interconnection of the open-loop system \eqref{eq:discrete_open_loop_small_signal} and control laws $\Delta\mathbf{u}_{\text{ref}j}$ in \eqref{eq:discrete_droop_compact} $\forall j\ne i$; $\boldsymbol{\xi}\sim \mathcal{N}(0,R)$ and $\boldsymbol{\gamma}\sim \mathcal{N}(0,V)$ are process and measurement noise respectively, and matrix $V$ is positive definite. Note that the system \eqref{eq:minimal_realization} is both controllable and observable. Besides, matrix $C_{\text{m}i}B_{\text{refm}i}$ is assumed to be full row rank.

With inputs $\{\Delta \mathbf u_{\text{ref}i},\Delta \mathbf u_{\text{L}}\}$ and outputs $\Delta \mathbf{y}_i$, the internal states $\Delta\mathbf{x}_i[k|k]$ can be estimated by the Kalman filter \cite{8345676}:
\begin{subequations} \label{eq:Kalman_filter}
	\begin{align}
		&\begin{aligned}
			\Delta\mathbf{x}_i[&k+1|k]=A_{\text{m}i}(I-GC_{\text{m}i})\Delta\mathbf{x}_i[k|k-1]+\\
			&B_{\text{refm}i}\Delta \mathbf u_{\text{ref}i}[k]	
			+ B_{\text{Lm}i}\Delta \mathbf u_{\text{L}}[k] + A_{\text{m}i}G \Delta \mathbf{y}_i[k],
		\end{aligned}\\
		& \Delta\mathbf{x}_i[k|k] = (I-GC_{\text{m}i}) \Delta\mathbf{x}_i[k|k-1]+G \Delta \mathbf{y}_i[k],
	\end{align}
\end{subequations}
where Kalman filtering gain $G=PC_{\text{m}i}^{\top}(C_{\text{m}i}PC_{\text{m}i}^{\top}+V)^{-1}$, whence $P$ is the positive definite solution to the Algebraic Riccati Equation \cite{8345676}.

Based on the above notations, we next present the following two matrices for detecting cyber attacks: given the reported measurement samples in a time window $T_0$,
\begin{subequations}
	\begin{align}
		& \begin{aligned}
			M&(T_0):=\frac{1}{T_0}\sum_k^{T_0}\{\Delta \mathbf{x}_i[k|k] - A_{\text{m}i}\Delta \mathbf{x}_i[k-1|k-1]-\\
			&B_{\text{refm}i}\mathbf{h}_{\text{d}i}(\Delta \mathbf{z}_i^{k-1})
			-B_{\text{refm}i}\mathbf{e}[k-1]-B_{\text{Lm}i}\Delta \mathbf{u}_{\text{L}}[k-1]\}\\
			&\{\Delta \mathbf{x}_i[k|k] - A_{\text{m}i}\Delta \mathbf{x}_i[k-1|k-1]-B_{\text{refm}i}\mathbf{h}_{\text{d}i}(\Delta \mathbf{z}_i^{k-1})\\
			&-B_{\text{refm}i}\mathbf{e}[k-1]-B_{\text{Lm}i}\Delta \mathbf{u}_{\text{L}}[k-1]\}^{\top}-GWG^{\top},
		\end{aligned}\\
		&\begin{aligned}
			&N(T_0):=\frac{1}{T_0} \sum_{k=1}^{T_0}\mathbf{e}[k-1]\{\Delta \mathbf{x}_i[k|k] - A_{\text{m}i}\Delta \mathbf{x}_i[k-1|k-1]\\
			&-B_{\text{refm}i}\mathbf{h}_{\text{d}i}(\Delta \mathbf{z}_i^{k-1})-B_{\text{refm}i}\mathbf{e}[k-1]-B_{\text{Lm}i}\Delta \mathbf{u}_{\text{L}}[k-1]\},
		\end{aligned}
	\end{align}
\end{subequations}
where $W:=C_{\text{m}i}PC_{\text{m}i}^{\top}+V$. The indicator matrices $M$ and $N$ are derived from the two tests of sensor veracity in \cite{7945354}. If the measurements $\Delta\mathbf{z}_i^k$ streaming to droop controllers at DGU $i$ is the same with the actual measurements $\Delta\mathbf{y}_i^k$, i.e., $\Delta\mathbf{z}_i^k\equiv \Delta\mathbf{y}_i^k$, all entries of both matrices $M$ and $N$ should be close to zeros with a sufficient large time window $T_0$; otherwise, large entries are expected to appear in either $M$ or $N$ or both.

A threshold test for detecting cyber attacks on a renewable-rich microgrid can be designed based on matrices $M$ and $N$. Define the following scalar attack indicators:
\begin{equation}
	\chi_1 :=|\trace({M})|, \quad \chi_2:= \norm{N}_1,
\end{equation}
where $\trace{(\cdot)}$ is the trace operator; and $\norm{N}_1$ is the absolute sum of all elements of $N$. Given a time window $T_0$ of samples fed into droop controllers, if attacks happens in the time window, then $\chi_1\ge\chi_1^*$ or $\chi_2 \ge \chi_2^*$ or both, where the thresholds $\chi_1^*$ and $\chi_2^*$ can be selected based on a subroutine proposed in \cite{8345676}.

\section{Case Study} 
\label{sec:case_study}
In this section, the attack detection method based on watermarking is tested in a Texas A\&M software testbed of microgrid. Firstly, the test microgrid is briefly described. Then, the cyber vulnerability of such a microgrid is identified by launching a destabilization attack. Finally, as will be shown, the cyber attack is detected timely by using the watermarking-based approach.

\subsection{Test System Description} 
\label{sub:system_description}
\begin{figure}[htb]
	\centering
	\includegraphics[width = 3.5in]{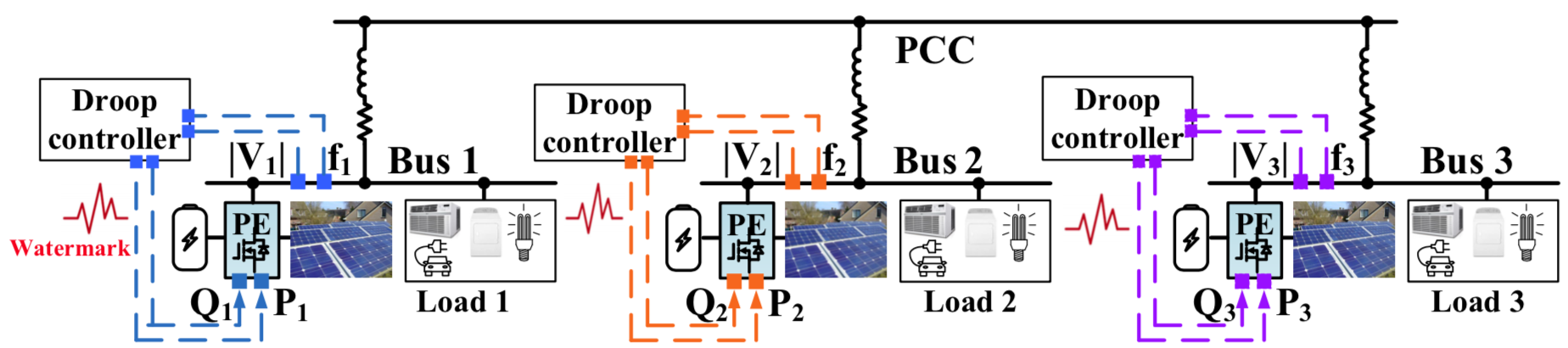}
	\caption{A renewable-rich microgrid.}
	\label{fig:microgrid}
\end{figure}

Fig. \ref{fig:microgrid} shows a four-bus, 220V microgrid operated in the islanded mode. The bus representing the point of coupling (PCC) is connected with the other three buses, each of which contains a load and a DGU, i.e. solar PV panels and energy storage. The distributed generations are interfaced to the grid through inverters. Operated in the islanded mode, this microgrid relies on the control of PE inverters to regulate its frequency and voltage. The test system parameters are summarized in Table \ref{tab:system_para}. As the test system includes all essential elements in a larger microgrid, i.e., DGUs equipped with control apparatus, AC network and loads, we leverage such a system to \emph{preliminarily} validate the efficacy of the proposed algorithm.

\begin{table}[htb]
	\caption{Summary of System Parameters}
	\label{tab:system_para}
	\centering

	\begin{tabular}{c|c|c|c}
	\hline

	\hline
	\textbf{Parameters} & \textbf{Values} & \textbf{Parameters} & \textbf{Values}\\
	\hline
	$R_{1-\text{PCC}}(\Omega)$& $0.23$& $V_{\text{dc}i}(V),i = 1,2,3$& $400$\\
	$X_{1-\text{PCC}}(\Omega)$& $0.10$& $R$ of VSC$i$ (m$\Omega$), $i=1,2,3$&$20.25$\\
	$R_{2-\text{PCC}}(\Omega)$& $0.30$& $X$ of VSC$i$ (m$\Omega$), $i=1,2,3$&$0.5$\\
	$X_{2-\text{PCC}}(\Omega)$& $0.35$& $K_{\text{p}1i}, i = 1,2,3$&$0.008$\\
	$R_{3-\text{PCC}}(\Omega)$& $0.35$& $K_{\text{i}1i}, i = 1,2,3$&$0.8$\\
	$X_{3-\text{PCC}}(\Omega)$& $0.58$& $K_{\text{p}2i}, i = 1,2,3$&$0.008$\\
	$T_{\text{f}i}(s),i = 1,2,3$& $0.02$& $K_{\text{i}2i}, i = 1,2,3$&$0.8$\\
	$T_{\text{V}i}(s),i = 1,2,3$& $0.02$& $\omega_{\text{n}}$(rad/s)&377\\
	$\alpha_{\text{p}_i},i = 1,2,3$&$-0.001$&$\alpha_{\text{q}_i},i = 1,2,3$&$-15.17$\\
	$\beta_{\text{p}_i},i = 1,2,3$&$-0.011$&$\beta_{\text{q}_i},i = 1,2,3$&$-769.44$\\
	
	\hline

	\hline
	\end{tabular}
\end{table}


\subsection{Impact of Cyber Attacks on Renewable-Rich Microgrid} 
\label{sub:impact_of_cyber_attacks_on_microgrid_with_deep_renewable}
\begin{figure}[ht]
		\centering
		\subfloat[]{\includegraphics[width=1.55in]{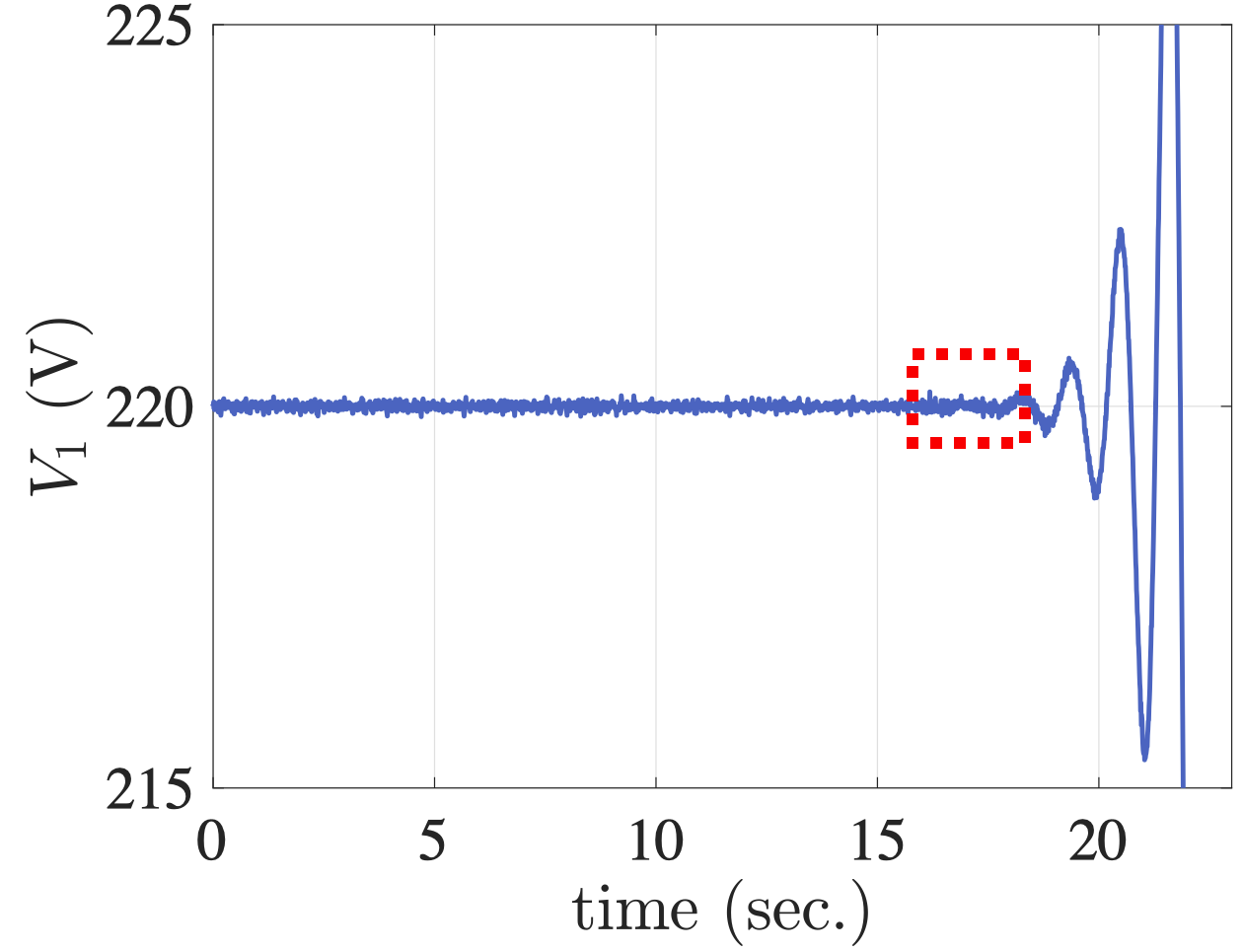}
		}
		\hfil
		\subfloat[]{\includegraphics[width=1.55in]{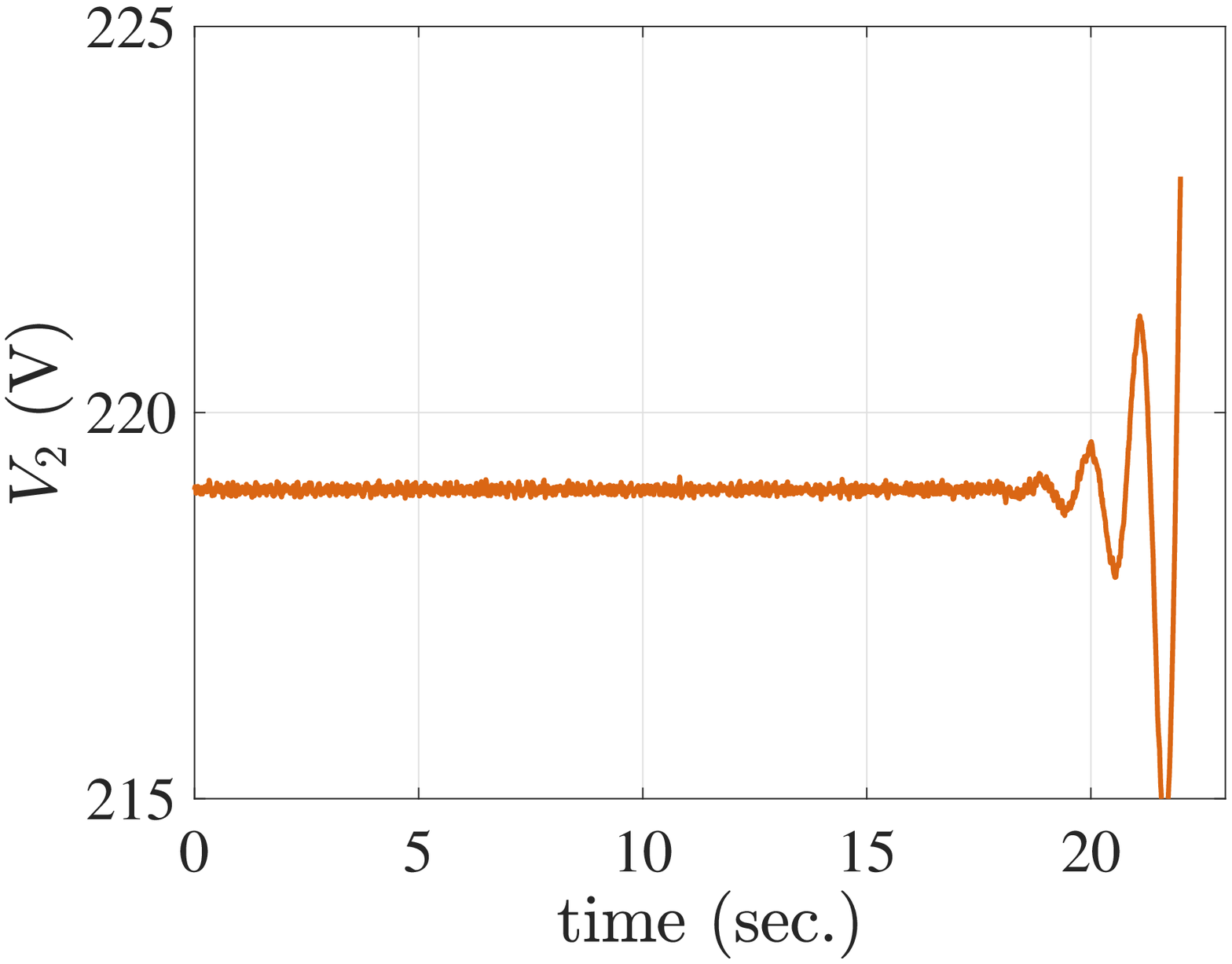}
		}
		\hfil
		\subfloat[]{\includegraphics[width=1.55in]{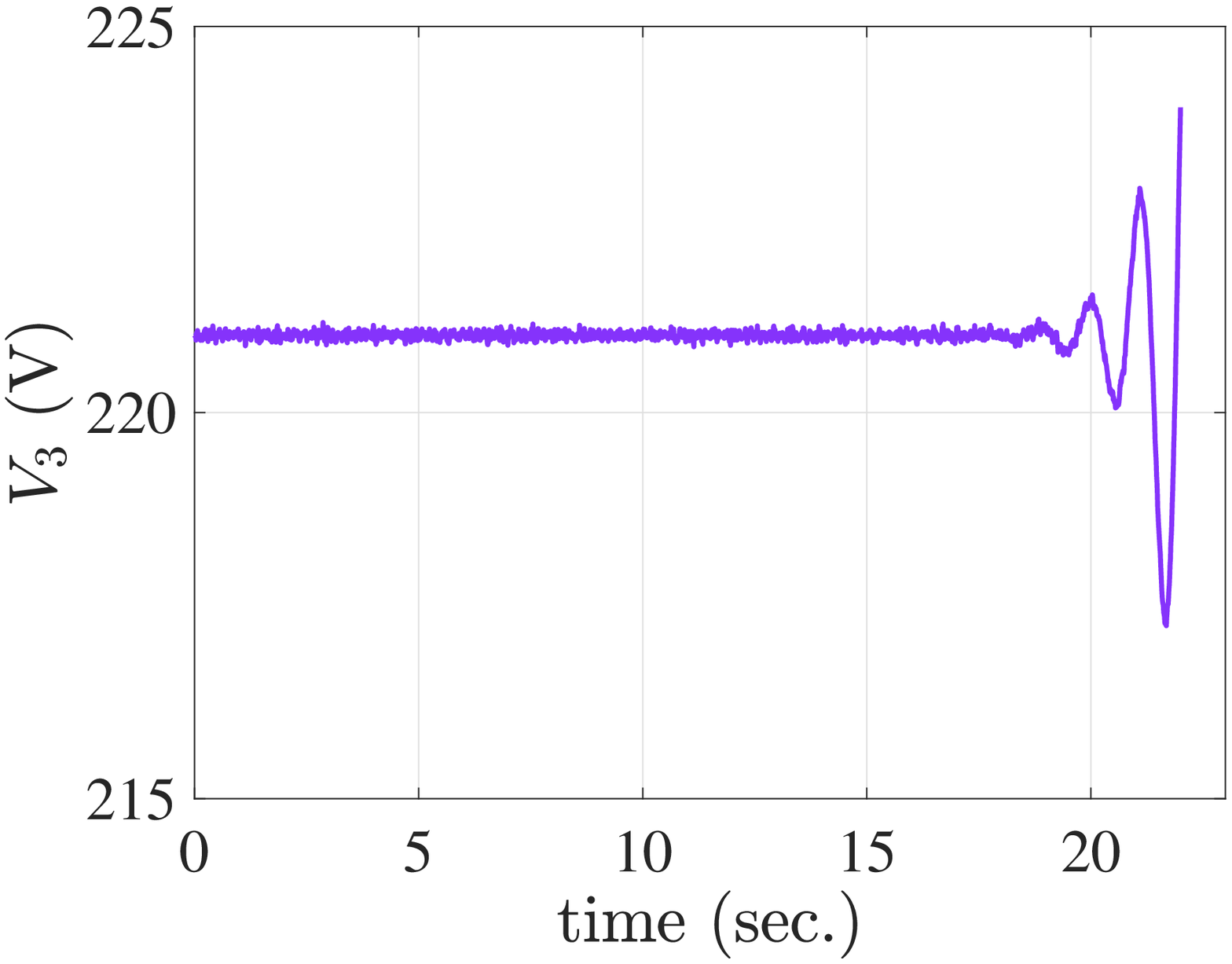}
		}
		\hfil
		\hfil
		\subfloat[]{\includegraphics[width=1.65in]{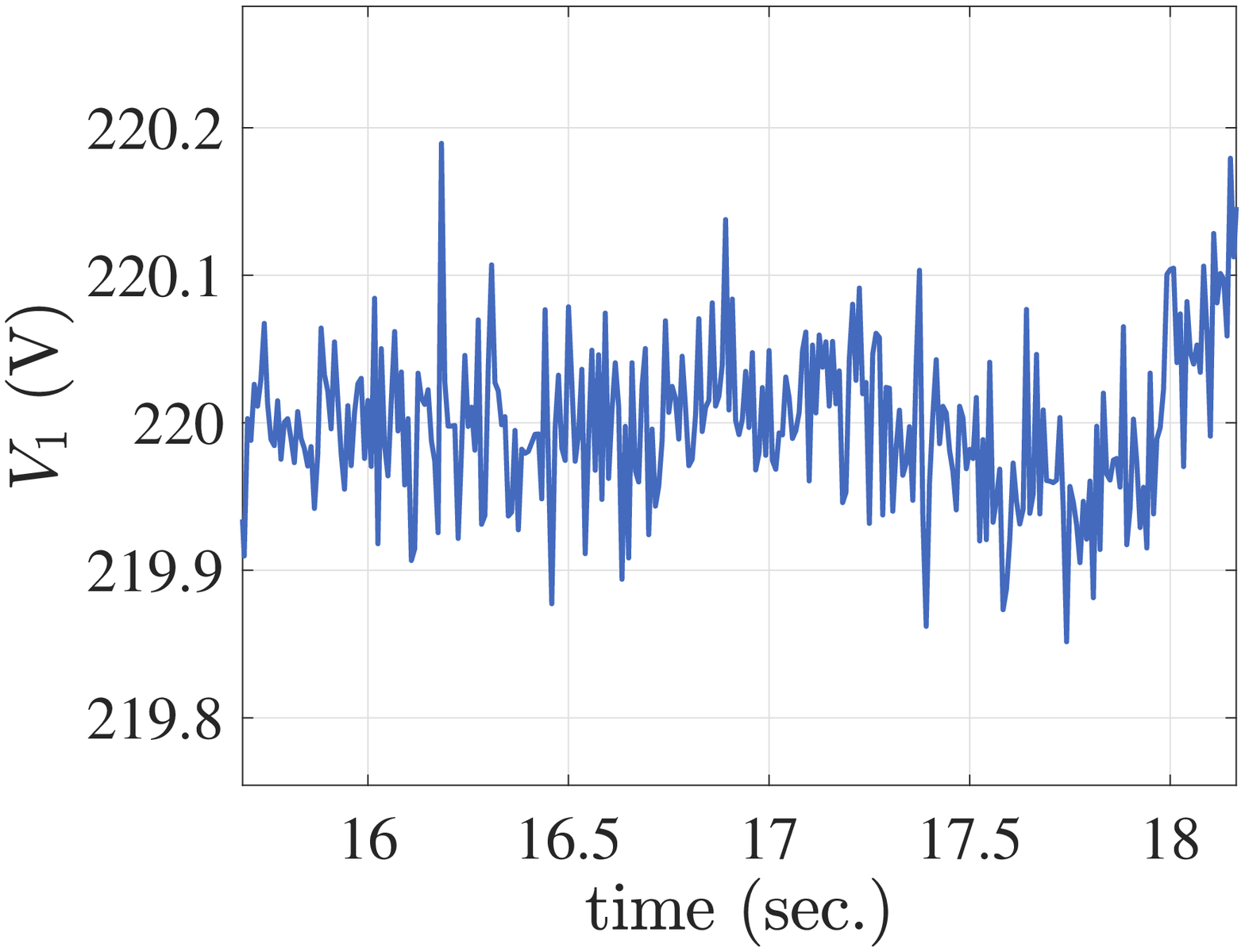}
		}
		\caption{Voltage magnitude at Bus $1$ (a), Bus $2$ (b), and Bus $3$ (c); the zoomed-in voltage magnitude at Bus $1$ around the $16$-th second (d).} \label{fig:bus_voltage} 
	\end{figure}

Adversaries can comprise the microgrid by manipulating measurements fed into one of the three droop controllers, say $\Delta V_1$ the voltage magnitude deviation at Bus $1$. Instead of reporting the actual $\Delta V_1$, the malicious sensors at Bus $1$ sends a ``filtered'' version of the measurement $\Delta V_1'$ to the droop controller. The reported $\Delta V_1'$ can be obtained by passing $\Delta V_1$ through a malicious filter and adding random noise to the filtered signal, i.e.,
\begin{equation}
	\Delta V_1'[k] =\mathcal{M}_0\left[\Delta V_1[k]\right]+\mu[k],
	\label{eq:malicious_filter}
\end{equation}
where the malicious filter $\mathcal{M}_0 = \frac{0.0012z+0.0008}{z^2-1.285z+0.2865}$ and $z$ is the $z$-transform operator; and $\mu\sim \mathcal{N}(0, 5\times 10^{-7})$. The first term of \eqref{eq:malicious_filter} denotes $m_0[k]*\Delta V_1[k]$, whence $m_0[k]$ is the inverse of $\mathcal{M}_0$, and ``$*$'' is the convolution operation. Such an attack is launched at the $16$-th second. As shown in Fig. \ref{fig:bus_voltage}, the attack incurs a system-level instability, although only the voltage magnitude at Bus $1$ suffers this cyber attack. Such an attack is termed as a \emph{destabilization attack} in \cite{8345676}. Fig. \ref{fig:bus_voltage}-d is a zoomed-in version of the voltage measurement at Bus 1 around the $16$-th second (the portion in the red-dash box in Fig. \ref{fig:bus_voltage}-a), i.e. the time instant when the attack is initiated. It is challenging to eyeball when the attack is initiated in Figure \ref{fig:bus_voltage}-d, as the voltage magnitude does not change much between the $16$-th second and the $18$-th second.

\subsection{Performance of the Detection Algorithm} 
\label{sub:results}
\begin{figure}[ht]
		\centering
		\subfloat[]{\includegraphics[width=1.6in]{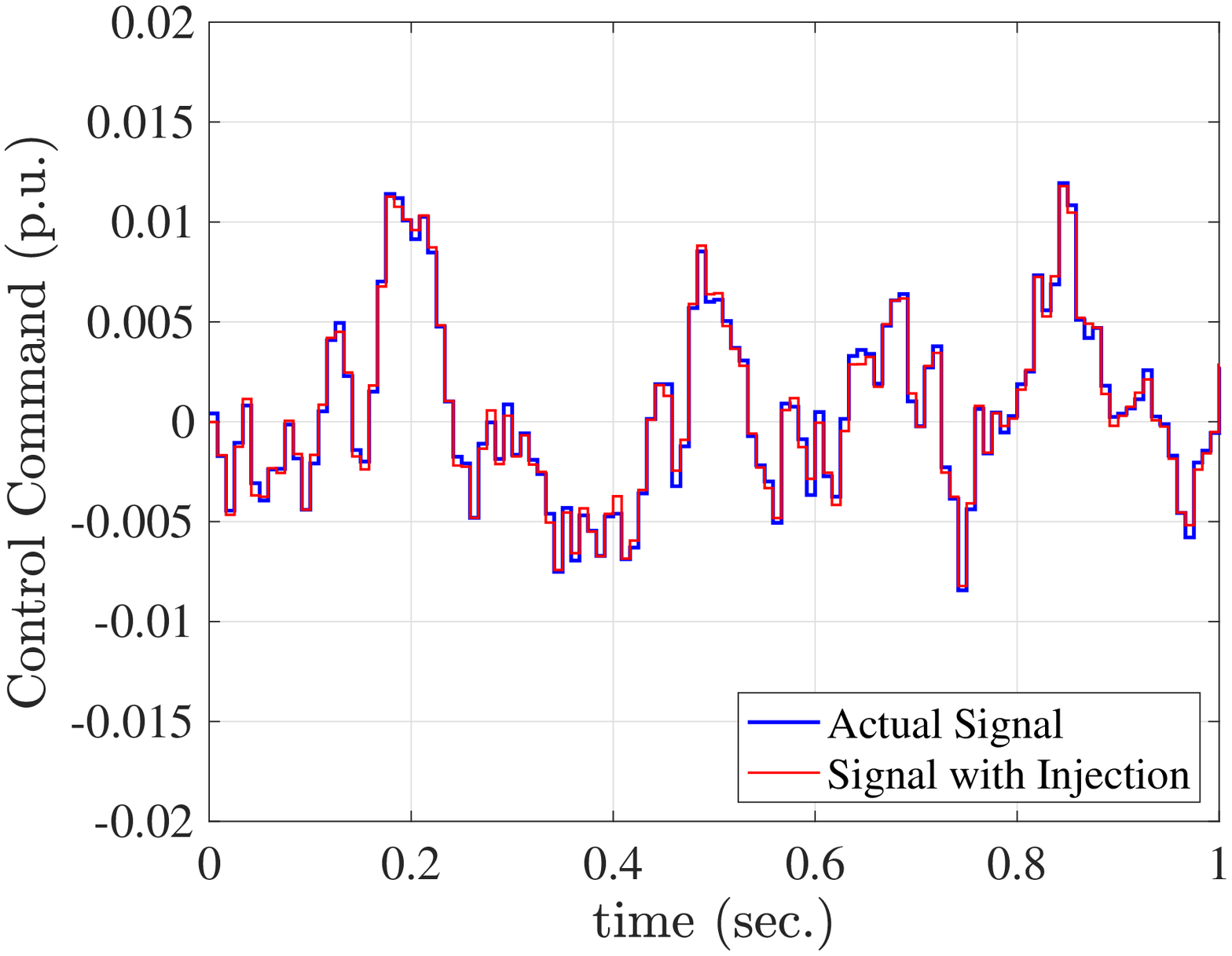}		}
		\hfil
		\subfloat[]{\includegraphics[width=1.6in]{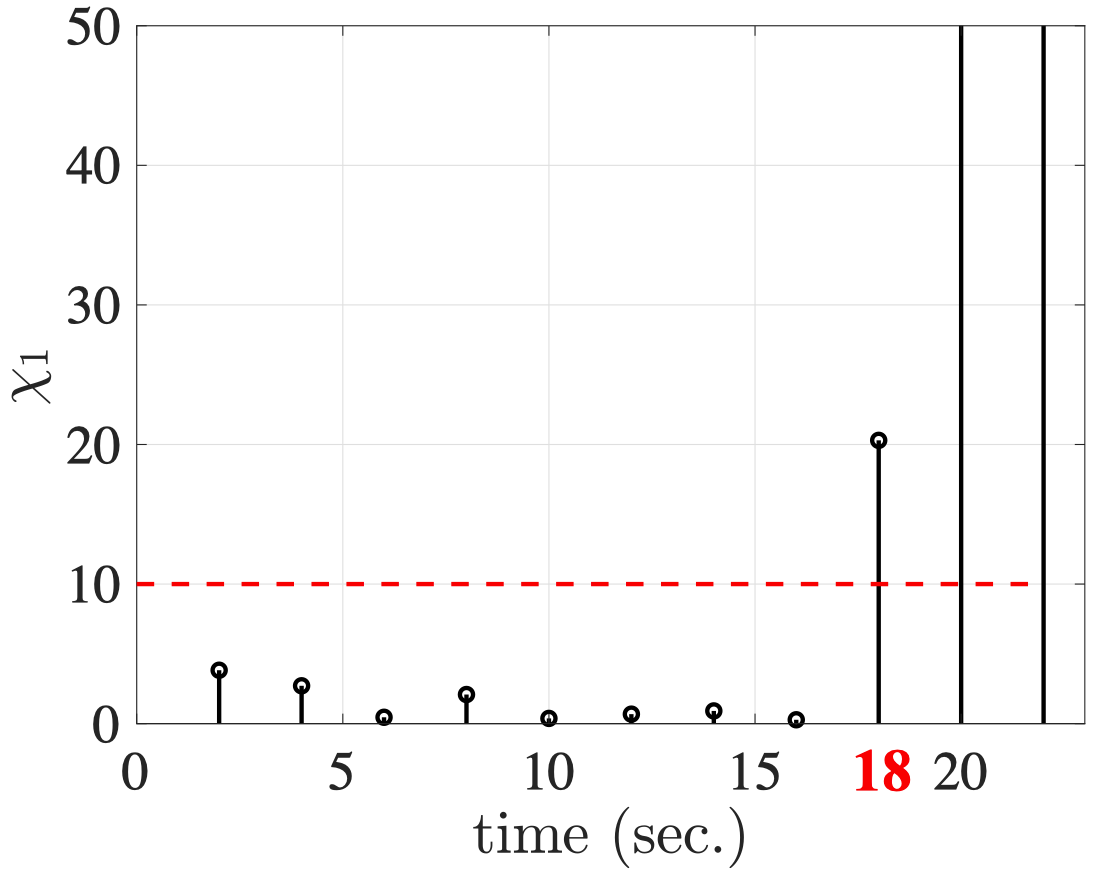}
		}
		\hfil
		\caption{(a) Impact of the watermark signal on the droop control command; (b) the evolution of the attack indicator $\chi_1$ under destabilization attack: the red dash line denotes a threshold.}
		\label{fig:detection}
	\end{figure}

In this subsection, we apply the dynamic watermarking-based algorithm to detect the attack introduced in Section \ref{sub:impact_of_cyber_attacks_on_microgrid_with_deep_renewable}. The parameters of the simulation are as follows: $\nu_e=10^{-7}$, $\chi_1^* = 10$, $T_0 = 2$ (second), and $\Delta T_{\text{s}}=0.0083$ (second). Fig. \ref{fig:detection}-a shows the actual control command without the watermark signal (blue) and the control command with watermark signal (red). These two signals are almost overlap with each other, indicating that the watermark signal does not compromise the performance of the droop controller. Fig. \ref{fig:detection}-b illustrates the evolution of the cyber attack indicator $\chi_1$: there are consecutive spikes exceeding the threshold $\chi_1^*$ after the $18$-th second, suggesting a cyber attack happens between the $16$-th second and $18$-th second. Note that it is theoretically guaranteed that, besides destabilization attack, the proposed method can detect any malicious distortion of distributed measurements fed into droop controllers \cite{7945354}.



\section{Conclusion} 
\label{sec:conclusion}
In this paper, we present the first demonstration of a dynamic watermarking approach to detecting cyber attacks in renewable-rich microgrids. A destabilization attack is adopted to expose cyber vulnerabilities of a renewable-rich microgrid. A cyber attack detection mechanism based on the dynamic watermarking is proposed to detect any possible cyber attacks on the measurements fed into droop controllers. The proposed mechanism is easily implementable, as no hardware upgrade is required, and it has theoretically provable performance in term of detecting cyber attacks. The efficacy of the proposed mechanism is tested and validated in a renewable-rich microgrid. Future work will extend the proposed mechanism to secondary and tertiary control of microgrids, and implement the mechanism in real-world microgrids.

\bibliographystyle{IEEEtran}
\bibliography{C4ref.bib}

\begin{thebibliography}{10}
\providecommand{\url}[1]{#1}
\csname url@samestyle\endcsname
\providecommand{\newblock}{\relax}
\providecommand{\bibinfo}[2]{#2}
\providecommand{\BIBentrySTDinterwordspacing}{\spaceskip=0pt\relax}
\providecommand{\BIBentryALTinterwordstretchfactor}{4}
\providecommand{\BIBentryALTinterwordspacing}{\spaceskip=\fontdimen2\font plus
\BIBentryALTinterwordstretchfactor\fontdimen3\font minus
  \fontdimen4\font\relax}
\providecommand{\BIBforeignlanguage}[2]{{%
\expandafter\ifx\csname l@#1\endcsname\relax
\typeout{** WARNING: IEEEtran.bst: No hyphenation pattern has been}%
\typeout{** loaded for the language `#1'. Using the pattern for}%
\typeout{** the default language instead.}%
\else
\language=\csname l@#1\endcsname
\fi
#2}}
\providecommand{\BIBdecl}{\relax}
\BIBdecl

\bibitem{4118327}
N.~{Pogaku} \emph{et~al.}, ``Modeling, analysis and testing of autonomous
  operation of an inverter-based microgrid,'' \emph{IEEE Trans. on Power
  Electronics}, vol.~22, no.~2, 2007.

\bibitem{1626398}
J.~A.~P. {Lopes} \emph{et~al.}, ``Defining control strategies for microgrids
  islanded operation,'' \emph{IEEE Trans. on Power Systems}, vol.~21, no.~2,
  2006.

\bibitem{8598671}
J.~A. {Ramos-Ruiz} \emph{et~al.}, ``Peer-to-peer energy transaction in
  microgrids with power electronics enabled angle droop control,'' in
  \emph{2018 eGrid}.

\bibitem{8345676}
T.~{Huang} \emph{et~al.}, ``An online detection framework for cyber attacks on
  automatic generation control,'' \emph{IEEE Trans. on Power Systems}, 2018.

\bibitem{Liu:2009:FDI:1653662.1653666}
Y.~Liu \emph{et~al.}, ``False data injection attacks against state estimation
  in electric power grids,'' in \emph{CCS 2009}.\hskip 1em plus 0.5em minus
  0.4em\relax ACM, 2009, pp. 21--32.

\bibitem{8260848}
H.~{Zhang} \emph{et~al.}, ``Distributed load sharing under false data injection
  attack in an inverter-based microgrid,'' \emph{IEEE Trans. on Industrial
  Electronics}, vol.~66, no.~2, pp. 1543--1551, Feb 2019.

\bibitem{7738534}
B.~Satchidanandan and P.~R. Kumar, ``Dynamic watermarking: Active defense of
  networked cyber physical systems,'' \emph{Proceedings of the IEEE}, vol. 105,
  no.~2, pp. 219--240, Feb 2017.

\bibitem{7945354}
------, ``On minimal tests of sensor veracity for dynamic watermarking-based
  defense of cyber-physical systems,'' in \emph{COMSNETS}, 2017.

\bibitem{kundur1994power}
P.~Kundur, N.~J. Balu, and M.~G. Lauby, \emph{Power system stability and
  control}.\hskip 1em plus 0.5em minus 0.4em\relax McGraw-hill New York, 1994,
  vol.~7.

\bibitem{8116598}
T.~Huang, M.~Wu, and L.~Xie, ``Prioritization of {PMU} location and signal
  selection for monitoring critical power system oscillations,'' \emph{IEEE
  Transactions on Power Systems}, 2017.

\bibitem{8362302}
T.~{Huang} \emph{et~al.}, ``Localization of forced oscillations in the power
  grid under resonance conditions,'' in \emph{CISS}, March 2018, pp. 1--5.

\bibitem{huang2018forced}
------, ``Forced oscillation localization in electric power systems under
  resonance conditions,'' \emph{arXiv preprint arXiv:1812.06363}, 2018.

\end{thebibliography}
\end{document}